\documentclass[final,5p,times,twocolumn,nopreprintline]{elsarticle}
\usepackage{amsmath,slashed,booktabs,dsfont}
\usepackage{epsfig,dcolumn}
\usepackage{graphicx}
\usepackage{comment}
\usepackage[utf8]{inputenc}
\usepackage{amsfonts}
\usepackage{amssymb}
\usepackage[dvipsnames,x11names]{xcolor}
\graphicspath{{figures/}}
\usepackage{xspace}
\usepackage{newtxtext,newtxmath}
\usepackage{bm}
\usepackage{subfigure}
\usepackage[colorlinks,linkcolor=NavyBlue,citecolor=NavyBlue,urlcolor=NavyBlue]{hyperref}
\usepackage{orcidlink}
\usepackage{fancyhdr}
\usepackage{color}
\usepackage{balance}
\usepackage{multirow}
\usepackage{multicol}
\biboptions{sort&compress}

\fancypagestyle{firstpage}{%
	
	\lhead{}
	\rhead{\small IPARCOS-UCM-25-020}
}
\newcommand{\be}{\begin{equation}}
\newcommand{\ee}{\end{equation}}

\newcommand{\rea}{{\rm Re}\,}
\newcommand{\ima}{{\rm Im}\,}

\usepackage{bm} 
\renewcommand*{\vec}[1]{\ensuremath{\bm{\mathrm{#1}}}}

\allowdisplaybreaks[4]

\begin{document}

\begin{frontmatter}
  
  \title{The role of chiral symmetry and the non-ordinary $\kappa/K^*_0(700)$ nature in $\pi^\pm K_S$ femtoscopic correlations} 

\author[ific]{M.~Albaladejo\orcidlink{0000-0001-7340-9235}} 
\author[ucm]{A.~Canoa\orcidlink{0009-0003-3515-5029}} 
\author[ific]{J.~Nieves\orcidlink{0000-0002-2518-4606}} 
\author[ucm]{J.R.~Pel\'aez\orcidlink{0000-0003-0737-4681}}
\author[ugr]{E.~Ruiz Arriola\orcidlink{0000-0002-9570-2552}}
\author[ucm]{J.~Ruiz de Elvira\orcidlink{0000-0001-6089-5617}}


  \address[ific]{Instituto de F\'isica Corpuscular (centro mixto CSIC-UV),
    Institutos de Investigaci\'on de Paterna, Apartado 22085, 46071, Valencia, Spain.}  
  \address[ucm]{Departamento de F\'isica Te\'orica and IPARCOS, Facultad de Ciencias F\'isicas,
    Universidad Complutense de Madrid, 28040 Madrid, Spain.}
  \address[ugr]{Departamento de F\'isica At\'omica, Molecular y Nuclear and Instituto Carlos I de F\'isica Te\'orica y Computacional, Universidad de
    Granada, E-18071, Granada, Spain.}

\begin{abstract}
  We show that the use of realistic $\pi K$ interactions, obtained from a dispersive analysis of scattering data, as well as relativistic corrections, are essential to describe recently observed $\pi^\pm K_S$ femtoscopic correlations. We demonstrate that the spontaneous chiral symmetry breaking dynamics and the non-ordinary features of the $\kappa/K^*_0(700)$ resonance, together with large cancellations between isospin channels, produce a large suppression of $\pi^\pm K_S$ femtoscopic correlations compared to widely used models. Within an improved version of the standard on-shell factorization formalism, we illustrate that compensating for this interaction suppression leads to source radii smaller than 1 fm, contrary to usual expectations, as well as larger correlation strengths. The relation between these two parameters cannot be accommodated within naive models describing the nature of the resonances. This may raise concerns about the applicability of popular but too simple approaches for systems with light mesons. However, the correlation-suppression effects we demonstrate here will be relevant in any formalism, and substantial corrections may be expected for other femtoscopic systems involving light mesons.
\end{abstract}

\end{frontmatter}

\thispagestyle{firstpage}

\section{Introduction}

There is a growing interest in femtoscopic correlations between particle pairs produced in relativistic collisions at LHC. 
These are defined as the ratio between the momentum distribution of pairs produced in the same event to the distribution when each particle is produced in different events.
Originally, these correlations were devised to study the emission source. However, they have been recently used to gain information on the pair interaction, particularly profiting from the 
huge statistics of the ALICE experiment at CERN or STAR at BNL. This could provide information on interactions inaccessible by other means
(i.e., between unstable states~\cite{ALICE:2020mfd,ALICE:2019eol}) or improve our knowledge when existing data are poor or extracted indirectly.
Indeed, over the last few years, ALICE and STAR have been studying correlations between pions and kaons and the resonances that appear there~\cite{ALICE:2023eyl,ALICE:2023sjd,ALICE:2020mkb,Abelev:2006gu}.

Very recently, ALICE has measured $\pi^\pm K_S $ femtoscopic correlations with remarkable accuracy~\cite{ALICE:2023eyl}.
They also explored the theoretical implications with a simplified model, using some radical assumptions, which hints at interesting features about the long-debated nature of the $\kappa/K^*_0(700)$ resonance. These quite popular approximations include a non-relativistic elastic formalism, neglecting the isospin 3/2 component, and a Breit-Wigner (BW) description of the $\kappa/K^*_0(700)$. 
We believe the ALICE high-quality data and their theoretical exploration point to interesting features worth investigating. However, we will show here the relevance of two often neglected effects. First, relativistic corrections and, second, realistic dispersive analyses of $\pi K$ scattering data. In particular, the $\kappa/K^*_0(700)$ non-ordinary nature and its interplay with the QCD spontaneous chiral symmetry breaking lead to large interaction cancellations that could appear in other systems with pions.

Indeed, the ALICE data can be described within a relativistic approach using realistic $\pi K$ interactions. For simplicity and to compare with ALICE, we use an improved version of a popular factorization approach, but these effects are a must for any formalism. Within this formalism, the source radii turn out smaller and the correlation strengths larger than ALICE's. In addition, their relation cannot be described by simple constituent models.

\section{Formalism}

\begin{figure}
\includegraphics[width=0.48\textwidth]{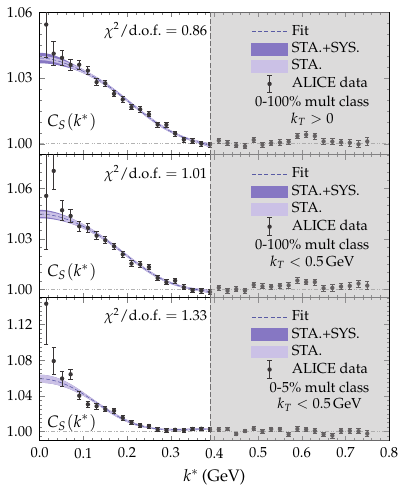}
 \caption{ALICE data on femtoscopic $S$-wave correlations for three different datasets. The blue lines and bands result from fits using our formalism with realistic $\pi K$ interactions and relativistic corrections. 
   The formalism presented here is valid in the elastic $\pi K$ regime, but we show data beyond the $\eta K$ threshold for completeness. 
 }\label{fig:ourresults} 
\end{figure}

In practice, the correlation function $C(k^*)\equiv A(k^*)/B(k^*)$, for a center-of-mass pair momentum $\vec{k^*}$, is measured by ALICE~\cite{ALICE:2023eyl} in $pp$ collisions.
Here, $A(k^*)$ is the distribution of pairs from the same event, and $B(k^*)$ is the reference distribution of pairs from mixed
events. 
Note that ALICE counts both $\pi^+ K_S$ and $\pi^- K_S$ pairs in $C(k^*)$, assuming their distributions are the same.
In the absence of $\pi K$ interactions $C(k^*)=1$. However, the raw $C(k^*)$ displays a non-flat baseline associated with soft parton fragmentation, or mini-jets, which is removed by normalizing it with a MonteCarlo (MC) simulation. This also removes the $K^*_0(892)$ vector resonance dominant peak, but not completely. The resulting ratio $C'(k^*)$ is 
attributed to actual femtoscopic correlations and then fit to:
\begin{equation}
C'(k^*) = \kappa\left[C_{\rm S}(k^*)+\epsilon \frac{dN_{BW}}{dm}\frac{dm}{dk^*}\right].
\label{eq:CS}
\end{equation}
Here, $\kappa$ is a normalization and $\epsilon$ 
is used to remove completely the $K^*_0(892)$ contribution, modeled with a simple BW distribution $dN_{BW}/dm=\Gamma/[(m-M)^2+\Gamma^2/4]$, whose mass $M$ and width $\Gamma$ are taken from the Review of Particle Physics~\cite{Workman:2022ynf}. Thus, in principle, $C_{\rm S}(k^*)$ corresponds to $S$-wave correlations, for which ALICE provides three datasets, I to III, with different multiplicity classes and $k_T$ cuts, shown in Fig.~\ref{fig:ourresults}.

From the theory side, the correlation function is customarily described within the Koonin-Pratt formalism~\cite{Koonin:1977fh,Yano:1978gk,Lisa:2005dd,Fabbietti:2020bfg} as $C(k^*)=\int d^{3}\vec{r}\,S(r) |\psi^{*}(\vec{k^*},\vec{r})|^2$. 
Here, $S(r)=\exp(-r^2/4R^2)/(4\pi R^2)^{3/2}$ is  the normalized source of radius $R$, and $\psi$ is the pair wave-function at a relative distance $\vec r$.
\begin{figure}
\includegraphics[width=0.4\textwidth]{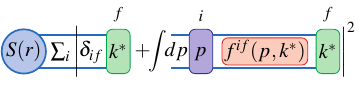}
\caption{Schematically, the final state $f$ is produced directly from the source $S$ or intermediate states $i$ (including also $f$) with virtual momentum $p$, through the half off-shell  $f^{if}$ amplitude. In factorization schemes, the $f^{ if}(p,k^*)$ amplitudes factorize on-shell out of the $p$ integral, i.e., $f^{ if}(k^*)$, leaving just the pair of propagators inside the $p$ integration. 
}\label{fig:LS}
\end{figure}

To explore the consequences of these data, ALICE describes $C_{\rm S}(k^*)$ in Eq.~\eqref{eq:CS} with the popular Lednick\'y–Lyuboshitz (LL) model ~\cite{Lednicky:1981su,Lednicky:2005af,Abelev:2006gu}, which is a non-relativistic factorization formalism.
In Fig.~\ref{fig:LS} we represent schematically the calculation of the pair wave function $\psi$ from the scattering amplitudes $f^{if}(p,k^*)$
using a Bethe-Salpeter approach (or Lippman-Schwinger if non-relativistic).
In factorization schemes, like the LL-model, the $f^{ if}(p,k^*)$ amplitudes factorize on-shell out of the $p$ integral, i.e., as $f^{if}(k^*)$, leaving just the propagators of the pair inside.
Below the $K\eta$ threshold, multiple meson states like $\pi\pi\pi K$ have not been observed to couple to $\pi K$. Hence, in that region, our final state $f=\pi^\pm K_S$ can be produced from any of the intermediate states $i=1,2,3=\pi^\pm K_S, \pi^\pm K_L, \pi^0 K^\pm$. This is a coupled-channel system; however, in the isospin basis, there are just two independent and {\it elastic} amplitudes $f_0^{(I)}$, with isospin $I$=1/2 and $I$=3/2. We consider equal sources for all $\pi K$ states, but this assumption can be easily generalized if necessary. Particularizing to the $S$-wave and assuming 
factorization, one arrives at:
\begin{align}
C_{\rm S}(k^*)=&1+ \frac{\lambda}{2}
\left(\frac{1}{3}\delta C^{(1/2)}(k^*)+\frac{2}{3}\delta C^{(3/2)}(k^*)\right),
\label{eq:ourCS}\\
\delta C^{(I)}(k^*)=&\frac{\rea f_0^{(I)}(k^*)}{R}I_1(s,R) -\frac{\ima f_0^{(I)}(k^*)}{R}I_2(s,R)\nonumber\\
&+\left|\frac{f_0^{(I)}(k^*)}{R}\right|^2 I_3(s,R)+ \Delta C^{(I)},
\label{eq:deltaC}\\
\Delta C^{(I)}(k^*)=&-\frac{|f^{(I)}_0(k^*)|^2}{2 R^3 \sqrt{\pi}} 
\frac{d\,\rea \big(1/f_0^{(I)}(k^*)\big)}{d (k^{*2})}. \label{eq:DeltaC}
\end{align}
Here, $s=( (m_{\pi}^2+k^{*2})^{1/2}+(m_{K}^2+k^{*2})^{1/2} )^2$ 
is the usual Mandelstam variable. The $\lambda$ parameter, called correlation strength, is included to measure the purity of genuine
pairs~\cite{Fabbietti:2020bfg,Lednicky:1981su,Lednicky:2005af,ALICE:2018ysd}. 

Incidentally, the 1/3 and 2/3 factors are the same as the Clebsch-Gordan coefficients in the $f_0^{\pi^+ K_S}=f_0^{(1/2)}/3+2f_0^{(3/2)}/3$ amplitude. 
The three $I_k(s)$ functions stem from the $p$ integrals in Fig.~\ref{fig:LS}, including the relativistic measure and the intermediate propagators. Namely,
$I_1=\rea{J_1}, I_2=\ima{J_1}$ and
\begin{align}
&J_1(s,R)= -16\pi R\sqrt{s}\!\!\int\!\!\! d^3\vec{r}\,S(r)\ j_0(k^*r)g(s,r),
\nonumber \\ 
&I_3(s,R)=64\pi^2 R^2 s\!\!\int\!\!\! d^{3}\vec{r}\, S(r)\left|g(s,r) \right|^2, \nonumber
\end{align}
where $j_0$ is a Bessel function (from the spherical $S$-wave) and
$$
g(s,r)=\int\!\!\!\frac{d^3\vec{p}}{(2\pi)^3}\frac{E_K+E_\pi}{2E_K E_\pi}
\frac{j_0(pr)}{s-(E_K+E_\pi)^2+i\epsilon}.
$$
Note the use of relativistic energies $E_a=\sqrt{m_a^2+p^2}$.

The $\Delta C^{(I)}$ terms account for the leading correction to the use of the asymptotic instead of the interacting wave function.

The usual $C_\text{LL}$ formula for one elastic channel~\cite{Lednicky:1981su,Lednicky:2005af,Abelev:2006gu} used by ALICE is recovered as follows:
Since they only consider the $I$=1/2 $\kappa/K_0^*(700)$ channel in $\pi^\pm K_S$, replace $C_S\to C_\text{LL}^{(1/2)}\equiv 1+(\lambda /2) \delta C_\text{LL}^{(1/2)}$. The $\text{LL}$ subscript means substituting the $I_k$ by $F_k$ functions obtained from the leading-order non-relativistic expansion of $g(s,r)$. Finally, 
the derivative in $\Delta C^{(1/2)}$ is replaced by its value at threshold~\cite{Lednicky:1981su,Albaladejo:2025kuv}, 
i.e., the effective range $d_0^{(1/2)}/2$.\footnote{Eq.~(8) in~\cite{ALICE:2023eyl} has a typo and should be multiplied by 2.}

\section{Results}

Theoretical implications were explored by ALICE in Ref.\,\cite{ALICE:2023eyl} using a toy $\pi^\pm K_S$ scattering amplitude, assuming it is elastic and dominated by a BW resonance, i.e., $f_0(k^*)=\gamma/(M_R^2-s-i\gamma k^*)$.
Fits to their I, II, and III datasets yielded, respectively: $M_R=833(15),\,804(14),\,765(13)\,\text{MeV}$ and $\gamma=890(25),\,801(33),\,714(^{+44}_{-39})\,\text{MeV}$. 
Despite these masses and couplings being incompatible, they were all identified with the same $I=1/2$ scalar $\kappa/K^*_0(700)$ meson. The source radius in these fits is about $1\,\text{fm}$ or larger, namely, $R=0.912(65),\,1.063(88),\,1.618(^{+163}_{-142})\,\text{fm}$.
In addition, the correlation strengths came out surprisingly small $\lambda=0.078(^{+13}_{-12}),\,0.111(22),\,0.274(^{+81}_{-59})$. We think that part of the $\lambda$ and $R$ variation could be due to using different, even incompatible, $\kappa/K^*_0(700)$ masses and widths on each fit.  Nevertheless, in~\cite{ALICE:2023eyl}, the observed increase of $\lambda$ with $R$, which is almost linear, was claimed to support a tetraquark nature.

We consider this ALICE theoretical exploration to be very suggestive. However, two relevant caveats must be addressed: relativistic corrections and realistic $\pi K$ interactions, the same in all three fits.  Both caveats are relevant no matter whether factorization is assumed or not. Our Eq.~\eqref{eq:CS} can deal with the two and provide an easy comparison with ALICE's treatment. 

\begin{figure}
\includegraphics[width=0.47\textwidth]{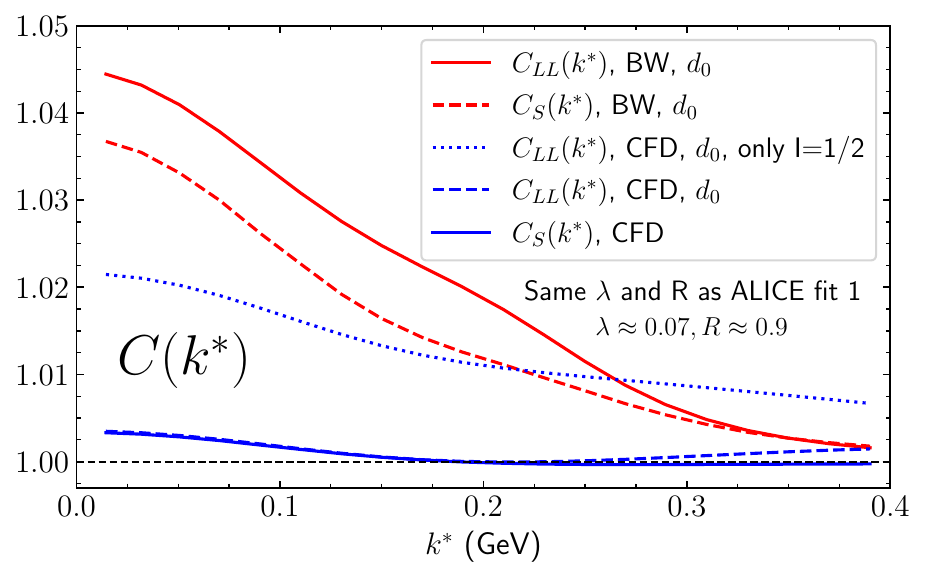}
\caption{The red curve is ALICE fit 1
 with the non-relativistic LL model and $I$=1/2 elastic BW $\pi^\pm K_S$ interaction only. In the red-dashed line, the LL formula is replaced by our relativistic equation. 
 The dotted blue line is the LL model with realistic $I$=1/2 $\pi K$ interactions (CFD,~\cite{Pelaez:2020gnd}) only. The continuous blue line is our relativistic model with realistic interactions for both $I$=1/2 and 3/2 but without 
 the derivative in $\delta C^{(I)}$ set to $d_0^{(I)}/2$ (as done for other lines).
}\label{fig:comparison} 
\end{figure}

Let us first discuss relativistic corrections, required since $m_\pi\!\!\simeq\!\!140\,$MeV, whereas the data reaches $k^*$ of hundreds of MeV. Moreover, the 
internal (virtual) momentum $p$ of intermediate pairs in Fig.~\ref{fig:LS}
is integrated from zero to infinity (but cut off by the source). Our Eq.~\eqref{eq:ourCS} uses a minimal modification~\cite{RuizdeElvira:2018hsv,Vidana:2023olz, Albaladejo:2024lam,Liu:2024nac} to the LL formalism, replacing the non-relativistic integration measure and propagators inside the integrals with their relativistic counterparts. These relativistic effects {\it decrease} the correlation. This is illustrated in Fig.~\ref{fig:comparison} by the difference between the ALICE result (continuous-red line) and the dashed-red curve, which includes relativistic effects while keeping, as ALICE does, only the $I$=1/2 toy BW interaction and all other parameters the same. 
The first surprise is that the correction does not vanish at $k^*=0$, but is quite large instead, i.e., $\sim20\%$. The reason is that besides $k^*$, there is also the internal (virtual) momentum $p$, which in Fig.~\ref{fig:LS} appears in two propagators and the integral measure. Since $p$ runs from zero to infinity, it yields significant relativistic contributions to the $I_i$ functions even if $k^*=0$. Second, it may seem counterintuitive that relativistic corrections seem larger at low momentum. Of course, in relative terms, relativistic contributions to the $I_i$ functions grow with momentum, but the overall huge suppression of the whole functions with $k^*$ masks these corrections in absolute terms above $k^*>0.3$ GeV. 

\begin{figure}
  \includegraphics[width=0.48\textwidth]{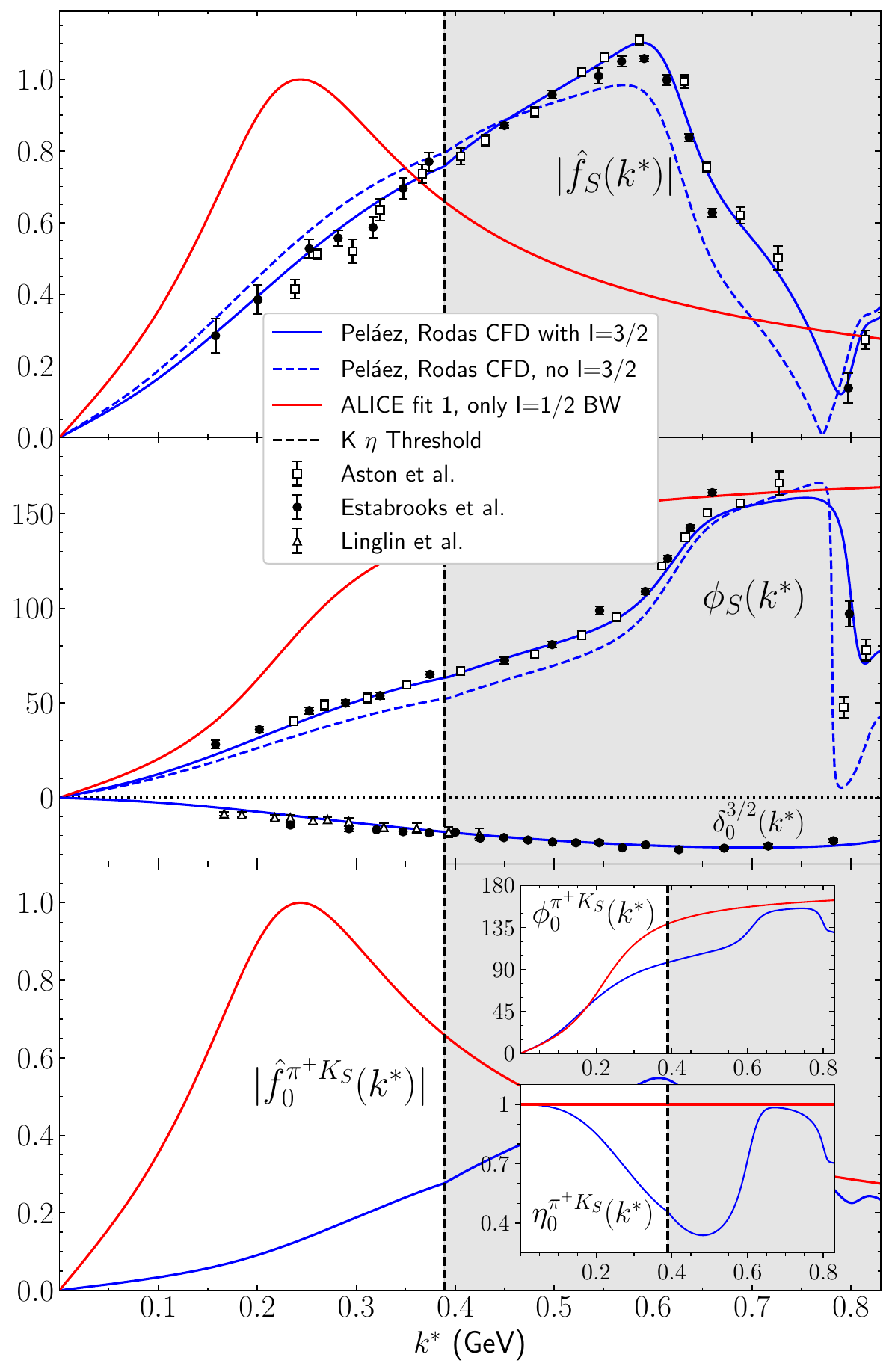}
\caption{$\pi K$ scalar scattering data \cite{Aston:1987ir,Estabrooks:1977xe, Linglin:1973ci}. Note the $\hat f=k^* f$ normalization.  
Top: $\vert \hat f_S\vert$. Center: $\phi_S$ and $\delta_0^{3/2}$. Red lines stand for the $I$=1/2 Breit-Wigner (BW) used by ALICE (fit 1, the other two are similar). Blue lines correspond to the CFD dispersive analysis of \cite{Pelaez:2020gnd}. Dashed lines are obtained by removing the $I$=3/2 contribution.
Bottom: modulus, phase, and elasticity
 of $\hat{f}^{\pi^+ K_S}_0$, relevant for femtoscopic $\pi^\pm K_S$ correlations, using CFD and the ALICE BW model.
}\label{fig:BWversusdata}
\end{figure}

Let us now discuss the effect of using realistic $\pi K$ interactions.
In the top and center panels of Fig.~\ref{fig:BWversusdata}, we show
the data on scalar $\pi K$ scattering.
Customarily, partial-wave data are given in terms of the
modulus and phase or elasticity and phase shift 
as $\hat{f}_0=\vert \hat f_0\vert \exp(i\phi)=
 (\eta\exp(2i\delta)-1)/2i$. Note the normalization $\hat f=k^* f$.
The repulsive $I$=3/2 channel has been measured directly in $\pi^+ K^+$ scattering~\cite{Linglin:1973ci,Estabrooks:1977xe} and, in practice, is elastic up to $k^*=0.8\,\text{GeV}$. Thus $\eta^{(3/2)}=1$ and it is fully determined by $\delta_{3/2}$. The other data come from~\cite{Aston:1987ir, Estabrooks:1977xe} for the modulus and phase of $\pi^+K^-$ scattering, i.e., $\hat f_S\equiv \hat f_0^{(1/2)}+\hat f_0^{(3/2)}/2=\vert \hat f_S\vert \exp{(i\phi_S)}$.
Continuous blue lines correspond to the dispersively constrained fits to data (CFD) from~\cite{Pelaez:2020gnd}. The dashed blue line shows that removing the $I=3/2$ has a small effect in $\pi^+K^-$ scattering, which is dominated by the attractive $I=1/2$. However, the  ALICE pure $I=1/2$ BW (red) is very far from the data, even qualitatively. 

The panels of Fig.~\ref{fig:BWversusdata}  illustrate two salient features of the $\pi K$ interaction and the $\kappa/K^*_0(700)$ meson, which are very relevant for correlations. The first one is that the realistic threshold behavior is much suppressed compared to an ordinary BW. This is because pions and kaons are pseudo-Goldstone bosons associated with the QCD SU(3) spontaneous chiral symmetry breaking. Actually, the BW threshold behavior is largely inconsistent with chiral perturbation theory (ChPT) results. In particular, the $I$=1/2 scattering length from the ALICE BW parameterization is $m_\pi a_0^{(1/2)}=$ $0.43(4),\,0.46(5),\,0.55(7)$ for fits I, II and III, respectively, whereas the ChPT two-loop result is $m_\pi a_0^{(1/2)}=0.224$~\cite{Bijnens:2014lea} and the CFD yields $m_\pi a_0^{(1/2)}=0.225(8)$~\cite{Pelaez:2020gnd}. 
Similarly, $m_{\pi}d_0^{(1/2)}=-1.83(5)$, $-2.03(8)$, $-2.28(^{+14}_{-12})$
for fits I, II, and III, respectively, at odds with the dispersive value $-3.8(4)$. The CFD $I$=3/2 effective range is huge: $m_{\pi}d_0^{(3/2)}=30(5)$.

The second feature is that, below $1.2\,\text{GeV}$, there is no evident $\kappa/K^*_0(700)$ ``Breit-Wigner peak", and the phase does not reach $90^{\rm o}$. The realistic $\kappa/K^*_0(700)$, despite dominating $S$-wave dynamics, 
does not saturate the unitarity bound $\vert \hat f(s)\vert\leqslant 1$ that holds below $K\eta$ threshold. 
In contrast, the elastic BW saturates this bound, largely overestimating $\vert f_S(s)\vert$ below $K\eta$ threshold, where ALICE finds its clearest correlation signal. 

Fig.~\ref{fig:comparison} shows the effect on $C_S$ of the chiral symmetry and the non-ordinary $\kappa/K^*_0(700)$ $I$=1/2 interaction suppression. When we replace in $C_\text{LL}^{(1/2)}$ the toy BW with realistic $I$=1/2 interactions, keeping the rest the same, correlations halve.

But there is a third effect. We already commented
that the isospin combination in Eq.~\eqref{eq:ourCS} is that of $f_0^{\pi^+ K}$. Hence, the $I$=3/2 component is enhanced by a factor of 2 instead of suppressed by a factor of 2 as in $\pi^+K^-$ scattering. Moreover, since the $I$=1/2 interaction is attractive and the $I$=3/2 is repulsive, this produces a huge cancellation, clearly seen in $\vert \hat f_0^{\pi^+ K_S}\vert$ in the bottom panel of Fig.~\ref{fig:BWversusdata}.
The modulus obtained with realistic interactions (blue) is much smaller 
than ALICE's result, due to the large cancellation between isospin components. Note in the insets that $\phi_0^{\pi^+ K_S}$ comes rather different beyond $k^*=0.2\,$GeV, but also that 
ALICE's toy model, with only an $I$=1/2 BW and neglecting $I=3/2$, makes $f_0^{\pi^+K_S}$ artificially elastic, i.e., $\eta_0^{\pi^+K_S}\equiv1$. 
The $\pi K$ $S$-wave cancellation at threshold $ a_{1/2}+2 a_{3/2} = 0 $
is a low-energy chiral theorem~\cite{Weinberg:1966kf}, which applies to a different isospin combination for other pion-hadron interactions.
Strong $\pi K$ phase-shift cancellations also extend to the lowest resonance region, as  first noted in~\cite{Venugopalan:1992hy} (in common with the
$\pi\pi$ case~\cite{Pelaez:2002xf,GomezNicola:2012uc,Broniowski:2015oha}).

Fig.~\ref{fig:comparison} shows that these three interaction-suppression effects, together with the relativistic suppression, 
lead to much smaller correlations when using realistic interactions inside Eq.~\eqref{eq:ourCS} (blue) than those of ALICE (red) for the same $\lambda$ and $R$.

\begin{table}\setlength{\tabcolsep}{0pt}
\caption{Parameters and $\chi^2/\text{d.o.f.}$ of the fits in Fig.\,\ref{fig:ourresults}. \label{tab:fitparam}}
\begin{tabular}{cccc}\hline
 & \parbox{2.1cm}{\vspace{1mm} \centering $0$-$100\%$ m.\,cl. \\ \centering $k_T>0$               \vspace{1mm}} & 
   \parbox{2.1cm}{\vspace{1mm} \centering $0$-$100\%$ m.\,cl. \\ \centering $k_T<0.5\,\text{GeV}$ \vspace{1mm}} &
   \parbox{2.1cm}{\vspace{1mm} \centering $0$-$5\%$   m.\,cl.   \\ \centering $k_T<0.5\,\text{GeV}$ \vspace{1mm}} \\ \hline\hline
$\chi^2/\text{d.o.f.}$ & $0.86$ & $1.01$ & $1.33$ \\
$R\,\text{(fm)}$ & $0.36(3)(3)$ & $0.41(3)(3)$ & $0.68(5)(3)$ \\
$\lambda$ & $0.19(2)(4)$ & $0.29(4)(5)$ & $0.80(14)(13)$ \\
$(N-1)\! \times\! 10^2$ & $0.80(8)(7)$ & $0.85(8)(6)$ & $0.97(6)(4)$ \\ \hline
\end{tabular}
\end{table}

Naively, Eq.~\eqref{eq:ourCS} suggests that the data fit
will compensate for the interaction overestimation by decreasing $\lambda$, increasing $R$, or both. 
Indeed, in Fig.~\ref{fig:ourresults}, we show our data fits using realistic interactions~\cite{Pelaez:2020gnd} within our relativistic formalism. We fit three parameters for each dataset: $\lambda$, $R$, and a normalization $N$, because the $\kappa$ normalization and $\epsilon$ in Eq.~\eqref{eq:CS} are not provided by ALICE but absorbed in their data. So we just multiply the $C_S$ data by $N$.
In contrast, ALICE fit those four 
parameters and a different $M_R$ and $\gamma$ for each dataset, whereas we use the same $\pi K$ interactions for all of them. The parameters of our fits are listed in Table~\ref{tab:fitparam}. 
The normalization barely changes by $\sim 1\%$. As expected, compared to the ALICE values given above, $\lambda$ becomes larger, and $R$ comes out surprisingly smaller than the usual $\gtrsim 1\,$fm
also obtained by ALICE for other hadronic correlations (see however~\cite{ALICE:2020mfd,ALICE:2022enj}). Note that fit 3, having the highest multiplicity class, is 
the most suitable for femtoscopic correlation studies and leads to relatively standard parameters. The very low radius for the other two fits casts doubts on the applicability of this simple formalism beyond illustrating the relevance of 
relativistic and realistic-interaction corrections. 

\begin{figure}
\centering
\includegraphics[width=0.5\textwidth]{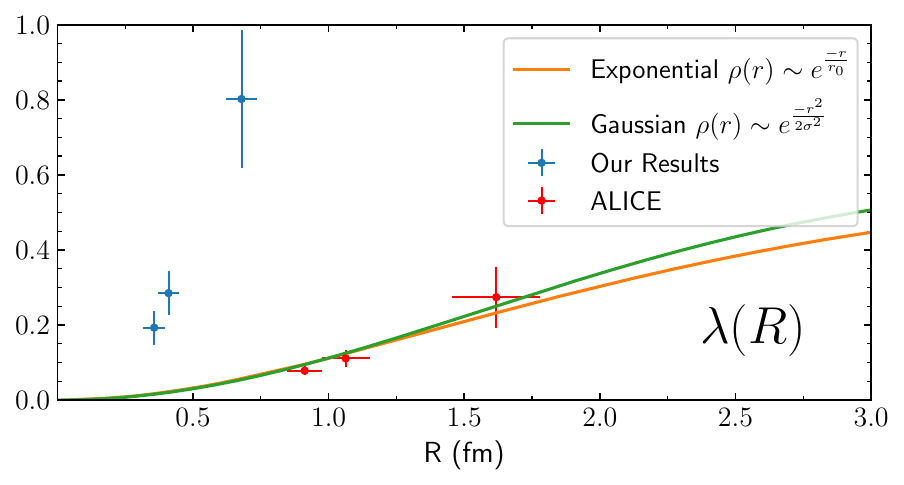}
\caption{\label{fig:R-lambda}$\lambda,R$ values found by ALICE in fits to their I, II, and III data sets. The curves correspond to fitting the $\lambda(R)$ dependence with the ALICE toy-model for tetraquarks described in the text. We find very different values when using realistic $\pi K$ interactions and incorporating relativistic corrections, which cannot be accommodated within the toy model for naive $q\bar q$ or tetraquark behaviors.}
\end{figure}

Finally, in Fig.~\ref{fig:R-lambda}, we show the $\lambda(R)$ dependence, which ALICE interprets using a toy model. They define either a Gaussian or exponential meson volume distribution $\rho(r)$ and a $\pi K$ ``overlap probability" $P=\int \rho(r)\rho(\vert \vec r-\vec R\vert) dV/\int \vert \rho(r)\vert^2 dV$. Note that P decreases with $R=|\vec R|$. Within this model, for an ordinary $q\bar q$ meson $\lambda$ decreases with $R$ as $\lambda=\lambda_0 a P$, whereas for tetraquarks it increases as $\lambda=\lambda_0 (1-aP)$. Based on previous references, they fix the ``$d\bar d$ annihilation efficiency'' $a=1$ and $\lambda_0=0.6$. Their $\lambda(R)$ values (in red) exclude a decreasing dependence. Their fits using this toy model are shown in Fig.~\ref{fig:R-lambda}, have $\chi^2/\text{d.o.f.}\lesssim0.4$, and yield $\sigma,\,r_0\simeq 1\,$fm as expected. Thus, they claim that their results suggest that the $\kappa/K_0^*(700)$ is a tetraquark.
However, the $\lambda,\,R$ values we find when using relativistic corrections and realistic $\pi K$ interactions (in blue) are remarkably different. Trying to fit them, we find a worse $\chi^2/\text{d.o.f.}\geqslant 1.5$, but with $\sigma,\, r_0\lesssim 0.26\,\text{fm}$. Therefore, the $\kappa/K^*_0(700)$ nature does not correspond to such a naive $q\bar q$ or tetraquark toy model. It might be worth studying whether the observed $\lambda(R)$ dependence can be reproduced within more elaborate descriptions consistent with chiral constraints and scattering data.

\section{Conclusions}

High-statistics femtoscopic correlations, such as those recently observed by ALICE for $\pi^\pm K_S$ pairs, may provide useful information on meson-meson interactions, particularly when data are scarce or inexistent. However, to calibrate these methods, it is important to understand first the cases when interactions are known. In this work, we have demonstrated the relevance of relativistic effects and of using realistic interactions when dealing with pions and kaons.
Indeed, relativistic corrections are relevant even at threshold
due to contributions from virtual propagators, not just external momenta.
Regarding meson-meson interactions, we have shown first that all isospin channels and their often large cancellations must be considered. Second, that they are overestimated when low-energy chiral dynamics and non-ordinary resonance behavior are ignored.
To ease the comparison with ALICE, we have used the popular formalism where scattering amplitudes factorize outside the intermediate-state momentum integrals. 
We think that the smaller-than-usual source radius obtained with this simplification should raise
concerns about its applicability to this and other similar systems and deserves further study.
However, any formalism will be affected both by relativistic effects and the overestimation of meson-meson amplitudes if naive resonant models are used.
Such an overestimation would still make correlations appear larger and consequently lead to artificially larger $R$ or smaller
$\lambda$, or both. 

In summary, the corrections we have worked out are not limited to the $\pi K$ system. Relativistic effects are important for femtoscopic correlations of any system involving pions and should not be neglected for kaons and etas.
Moreover, the QCD spontaneous chiral symmetry breaking threshold suppression and the non-ordinary behavior of the $\kappa/K^*_0(700)$ are also present in $S$-wave $\pi\pi$ interactions and the $\sigma/f_0(500)$~\cite{Pelaez:2021dak} as well in other light scalar resonances appearing in $\pi\pi$, $K\bar K$ or $\pi\eta$ interactions. Finally, large cancellations between different isospin scalar interactions are also common to other pion-hadron processes.

\section*{Acknowledgments}

 We thank O. V\'azquez Doce and T. Humanic for their comments and clarifications. This work is part of the Grants PID2022-136510NB-C31, PID2023-147458NB-C21, PID2020-112777GB-I00, CEX2023-001292-S, PID2020114767GB.I00 and PID2023.147072NB.I00 of MICIU/AEI, as well as CIPROM/2023/59 contract of the Generalitat Valenciana (GVA), FQM225 grant of Junta de Andaluc\'{\i}a, and the European Union’s Horizon 2020 research and innovation program under grant agreement No 824093 (STRONG2020). 
M.\,A.\,acknowledges financial support through GenT program by Generalitat Valencia (GVA) Grant CIDEGENT/2020/002, Ramón y Cajal program Grant RYC2022-038524-I of MICIU/AEI, and Atracción de Talento program by CSIC PIE 20245AT019.
A.\,C.\,is supported by the Spanish Comunidad de Madrid, fellowship CT85/23.
J.\,R.\,E.\, acknowledges financial support through the Ram\'on y Cajal program Grant RYC2019-027605-I of MICIU/AEI. %

\bibliographystyle{apsrev4-1}
\bibliography{kappafemtoshort}

\end{document}